\documentclass[runningheads]{llncs}
\usepackage[T1]{fontenc}
\usepackage{amssymb, amsmath}
\usepackage{graphicx,verbatim}
\usepackage{marvosym} 

\usepackage{bbding}

\DeclareRobustCommand{\equalcontrib}{\textsuperscript{\ensuremath{\dagger}}}
\DeclareRobustCommand{\corrauth}{\textsuperscript{\Envelope}}

\newcommand{\nonumfootnote}[1]{%
  \begingroup
  \renewcommand{\thefootnote}{}%
  \footnote{#1}%
  \addtocounter{footnote}{-1}%
  \endgroup
}
\usepackage{xurl}
\usepackage[colorlinks=true,citecolor=blue,linkcolor=blue,urlcolor=blue]{hyperref}

\begin{document}
%
\title{Defining Robust Ultrasound Quality Metrics \\ via an Ultrasound Foundation Model}
\titlerunning{Ultrasound Quality Metrics}
%


\author{Ziyang Huang\inst{1}\equalcontrib \and
Bingyan Li\inst{1}\equalcontrib \and
Chen Ma\inst{1}\equalcontrib \and
Tianyi Liu\inst{2,3} \and
Yihui Zhai\inst{2,3} \and
Hong Xu\inst{2,3} \and
Yi Guo\inst{1} \and
Zeju Li\inst{1}\corrauth \and
Yuanyuan Wang\inst{1}\corrauth}


\authorrunning{Huang et al.}

\institute{
College of Biomedical Engineering, Fudan University, Shanghai, China\\
\email{zejuli@fudan.edu.cn; yywang@fudan.edu.cn}
\and
Department of Nephrology, Children's Hospital of Fudan University,
National Children's Medical Center, Shanghai, China
\and
Shanghai Kidney Development and Pediatric Kidney Disease Research Center,
Shanghai, China
}

\maketitle

\begingroup
\renewcommand{\thefootnote}{}
\footnotetext{\textsuperscript{\ensuremath{\dagger}} Equal contribution; \textsuperscript{\Envelope} corresponding author.}
\endgroup

\begin{abstract}
Clinicians lack a principled framework to quantify diagnostic utility in ultrasound reconstructions.
Existing standards like PSNR and VGG-LPIPS are inadequate, failing to account for modality-specific physics or the structural nuances of acoustic imaging.
We close this gap with a TinyUSFM-based evaluation framework featuring two distinct metrics: TinyUSFM-uLPIPS, a full-reference perceptual distance based on multi-layer token relations, and TinyUSFM-NRQ, a deployable no-reference quality score utilizing clean-manifold modeling and worst-region aggregation to detect localized harmful artifacts. 
We demonstrate that the presented metrics have four unique advantages: 1) Task-linked quality, where TinyUSFM-uLPIPS achieves superior calibration with semantic task damage, accurately reflecting Dice-score drops in segmentation where VGG-based metrics fail; 2) Cross-organ comparability, maintaining stable scoring scales and consistent severity rankings across diverse anatomical sites and domain-shifted data; 3) PSNR-consistent sensitivity, with TinyUSFM-NRQ providing a reliable quality score without ground-truth images that remains consistent with traditional fidelity benchmarks (i.e. PSNR); and 4) Clinical utility, improving the prediction of expert preference from 47.2$\%$ to 72.8$\%$ accuracy and producing super-resolution reconstructions preferred by sonographers. By integrating these advantages into a unified assessment and optimization loop, this work establishes a modality-aligned standard that finally bridges the gap between algorithmic performance and diagnostic utility. Our code is available at \url{https://github.com/sextant-fable/US-Metrics}.
\end{abstract}

\section{Introduction}
Ultrasound is ubiquitous in clinical imaging due to its portability and safety. 
However, evaluating the perceptual and clinical usefulness of ultrasound reconstructions and generative outputs remains poorly served by commonly used image-quality metrics, especially across organs, devices, and acquisition settings.

Ultrasound appearance is shaped by modality-specific physics and sampling. Coherent interference produces speckle, specular responses are view dependent, attenuation and shadowing create structured dropouts, and scan conversion introduces anisotropic resolution and sampling artifacts \cite{bamber1986speckle,baad2017usartifacts,robinson1982scanconv}. 
Clinically meaningful failures are often localized: a small region of clutter or shadowing may erase a critical boundary even when global intensity changes slightly. In contrast, adjustments to global amplification or depth-dependent amplification alter pixel values across the image without degrading its diagnostic interpretability.
\begin{figure}[t]
\centering
\includegraphics[width=\textwidth]{graph1.png}
\caption{Overview of the TinyUSFM-based ultrasound quality evaluation framework, including metric motivation (left), proposed full-reference (FR) and no-reference (NR) metrics (middle), and target properties with shared-feature downstream utility (right).}
\label{fig:overview}
\end{figure}

Existing metrics are poorly calibrated for ultrasound due to three primary limitations:
1) Fidelity vs. Clinical Utility: Pixel-level measures (PSNR and SSIM~\cite{wang2004ssim}) prioritize global intensity agreement, making them overly sensitive to benign brightness shifts while failing to detect localized, clinically harmful artifacts.
2) Domain Mismatch: Natural-image perceptual metrics (VGG-LPIPS and FID~\cite{zhang2018perceptual,heusel2017ttur}) rely on photographic priors that misinterpret ultrasound-specific textures as noise and under-penalize structured occlusions.
3) Lack of Generalization: While generic no-reference IQA metrics such as BRISQUE~\cite{mittal2012brisque} and NIQE~\cite{mittal2013niqe}, and ultrasound-specific learned IQA/QA methods including fetal ultrasound IQA~\cite{wu2017fuiqa,zhang2021fetalqa_medicine}, expert-agnostic ultrasound IQA~\cite{raina2023us2qnet}, and KDE-based QA~\cite{kwon2023kdeuqa}, have been proposed, most existing approaches are tied to restricted clinical protocols, organ-specific assumptions, or task-specific labels. In contrast, our goal is to define both FR and NR metrics in a shared ultrasound foundation-model feature space and evaluate whether this space supports task-linked, cross-organ, and clinically aligned quality assessment.

As illustrated in Fig.~\ref{fig:overview}, we turn to USFM~\cite{jiao2024usfm} and its lightweight variant, TinyUSFM~\cite{ma2025tinyusfm} to develop a principled, ultrasound-native evaluation framework that bridges the gap between image-quality metrics and diagnostic utility. By leveraging the foundation model's feature space, we introduce two metrics—the full-reference (FR) TinyUSFM-uLPIPS and the no-reference (NR) TinyUSFM-NRQ—which offer superior calibration with clinical tasks and stable performance across diverse anatomical sites. We validate both metrics using task-based semantic anchors and clinician two-alternative forced choice (2AFC) studies. Furthermore, we employ the same feature space as a perceptual loss for super-resolution, establishing a validated pipeline for clinical image enhancement.

\section{Methodology}

\subsection{Ultrasound-native representations from TinyUSFM}
\label{sec:backg}
For an input image $x$, let $f_\ell(x)\in\mathbb{R}^{(1+T_\ell)\times C_\ell}$ denote the TinyUSFM output at layer $\ell$, where $T_\ell$ is the number of spatial patch tokens and $C_\ell$ is the channel dimension. After removing the global [CLS] token, we denote the remaining patch-token matrix by $F_\ell(x)\in\mathbb{R}^{T_\ell\times C_\ell}$. For TinyUSFM-NRQ, each layer is further summarized by average pooling its patch tokens to obtain $u_\ell(x)\in\mathbb{R}^{C_\ell}$, and the pooled features from layers $\mathcal{L}$ are concatenated and $L_2$-normalized to form a global descriptor $z(x)$. Thus, $F_\ell(x)$ is used for full-reference comparison, whereas $z(x)$ is used for no-reference quality modeling.

\subsection{TinyUSFM-uLPIPS: a full-reference quality metric}
Given a reference image $x$ and a test image $y$, TinyUSFM-uLPIPS compares their TinyUSFM patch-token features across layers $\mathcal{L}=\{3,5,7,11\}$, where $F_\ell(x),F_\ell(y)\in\mathbb{R}^{T_\ell\times C_\ell}$ denote the [CLS]-removed patch-token matrices at layer $\ell$. In all experiments, we use $\mathcal{L}=\{3,5,7,11\}$, $r=3$, and $\tau=20$.

Let $\hat{F}_\ell(x)$ and $\hat{F}_\ell(y)$ denote the row-wise $L_2$-normalized versions of $F_\ell(x)$ and $F_\ell(y)$. For each token location $q\in\{1,\dots,T_\ell\}$, let $\mathcal{N}_r(q)$ denote its local neighborhood with radius $r$, and let $A_{\ell,q}(x)\in\mathbb{R}^{|\mathcal{N}_r(q)|\times C_\ell}$ be the matrix formed by stacking the normalized token vectors in that neighborhood. We then define
\begin{equation}
d^{\mathrm{s}}_\ell(x,y)=\frac{1}{T_\ell}\sum_{q=1}^{T_\ell}\mathrm{MeanAbs}\!\left(\mathrm{Softmax}\!\left(\tau A_{\ell,q}(x)A_{\ell,q}(x)^\top\right)-\mathrm{Softmax}\!\left(\tau A_{\ell,q}(y)A_{\ell,q}(y)^\top\right)\right),
\end{equation}
where $\mathrm{Softmax}(\cdot)$ is applied row-wise and $\tau$ is a temperature parameter.

Channel co-activation statistics are compared using Gram matrices computed from the normalized token matrices \cite{gatys2015style}, with $G_\ell(x)=\frac{1}{T_\ell C_\ell}\hat{F}_\ell(x)^\top\hat{F}_\ell(x)$ and $d^{\mathrm{g}}_\ell(x,y)=\mathrm{MeanAbs}\!\left(G_\ell(x)-G_\ell(y)\right)$. The per-layer distance is $d_\ell(x,y)=d^{\mathrm{s}}_\ell(x,y)+d^{\mathrm{g}}_\ell(x,y)$, and the final TinyUSFM-uLPIPS score is
\begin{equation}
\mathrm{TinyUSFM\text{-}uLPIPS}(x,y)=\frac{1}{|\mathcal{L}|}\sum_{\ell\in\mathcal{L}} d_\ell(x,y).
\end{equation}

\subsection{TinyUSFM-NRQ: a no-reference quality metric}
TinyUSFM-NRQ scores a single ultrasound image without a reference by combining organ-aware clean-manifold likelihood with worst-region aggregation. This design models valid appearance variation across devices and settings while remaining sensitive to localized failures.

For an image $x$, we extract overlapping patches $\{x_i\}_{i=1}^{N_x}$ of size $224\times224$ with stride 112, where $N_x$ is the number of extracted patches. Each patch $x_i$ is encoded as $\hat z_i=z(x_i)$ using the global TinyUSFM descriptor from Sec.~\ref{sec:backg}. For each organ $o$, clean patch descriptors are projected to a PCA space of dimension $d=128$ and modeled by a diagonal-covariance Gaussian mixture model $p_o$ with $K=4$ components. If the organ label is available, patch $x_i$ is scored by $s_i(x)=\log p_o(\hat z_i)$; otherwise, we use a uniform mixture over all $O$ organ models, i.e., $s_i(x)=\log\!\left(\frac{1}{O}\sum_{o=1}^{O}p_o(\hat z_i)\right)$. To emphasize localized failures, we average the lowest-scoring patches, where $\kappa=\max\!\left(1,\mathrm{round}(0.15\,N_x)\right)$ and $\mathcal{W}_\kappa(x)$ denotes the index set of the $\kappa$ worst patches. We then define
\begin{equation}
\mathrm{TinyUSFM\text{-}NRQ}(x)=\frac{1}{\kappa}\sum_{i\in\mathcal{W}_\kappa(x)} s_i(x),
\end{equation}
where higher values indicate better quality.

\subsection{TinyUSFM features as a perceptual loss for image restoration}
\label{sec:percep}
For SR optimization, we use a differentiable token-distance loss in the same TinyUSFM feature space:
Let $\hat{F}_\ell(\cdot)$ denote the row-wise $L_2$-normalized [CLS]-removed token matrix at layer $\ell$. Given a reconstruction $\hat y$ and its target image $y$, we define
\begin{equation}
\mathcal{L}_{\mathrm{TinyPerc}}(\hat y,y)=\frac{1}{|\mathcal{L}|}\sum_{\ell\in\mathcal{L}}\frac{1}{T_\ell}\sum_{t=1}^{T_\ell}\left\|\hat{F}_{\ell,t}(\hat y)-\hat{F}_{\ell,t}(y)\right\|_2^2.
\end{equation}
For $512{\times}512$ images, the TinyUSFM perceptual loss is computed on sliding $224{\times}224$ windows and averaged across windows to avoid resizing artifacts.

\section{Experiments and Results}
\label{sec:experiments}

\subsection{Evaluation Protocol}
\label{sec:datasets_preproc}

\textbf{Task-Linked Quality (FR):} We assess metric ability to capture task-linked quality by using performance degradation of downstream models as an objective proxy for clinical utility. We systematically apply distortions~\nonumfootnote{The evaluation covers 8 distortions categorized into four primary groups: noise and texture corruption, blur and resolution loss, ultrasound-specific artifacts (such as shadowing, specular clipping, and missing scanlines), and geometric deformation.} using a binary search to ensure each reaches an identical PSNR-matched magnitude relative to the clean baseline. By evaluating a pre-trained model across these controlled degradations, we use the resulting drop in task performance (e.g., segmentation or classification accuracy) as a "ground truth" for how much a specific distortion actually interferes with diagnostic information. The rank correlation (Spearman’s $\rho$ and Kendall’s $\tau$) between TinyUSFM-uLPIPS and these accuracy drops serves as a primary indicator of whether the metric prioritizes task-relevant structural integrity over benign global fluctuations like intensity shifts.

\textbf{Cross-Organ Comparability (FR):} To evaluate cross-organ comparability, we assess the consistency of degradation rankings derived by TinyUSFM-uLPIPS across diverse anatomical sites using Kendall’s $W$, ensuring that the relative severity of distortions is penalized uniformly regardless of the organ. We further analyze score-scale stability using the interquartile range (IQR), which measures the statistical spread of the middle 50$\%$ of the data. A lower IQR indicates that the metric produces more consistent and predictable scores for a given quality level, suggesting that the numerical scale remains stable and is not skewed by anatomical variations.

\textbf{PSNR-Consistent Sensitivity (NR):} To evaluate our NR metric for TinyUSFM-NRQ, we utilize a ground-truth ranking protocol. We synthesize six variants of each clean image by applying distortions of known, increasing severity. Since the true quality hierarchy is established by design, we assess the metric’s effectiveness by measuring how closely its predicted scores correlate with these pre-defined ranks. High correlation indicates that the framework can accurately perceive diagnostic degradation without requiring a pristine reference image.

\textbf{Clinician Utility (NR):} To validate the metric's practical utility, we conducted a blinded 2AFC study. Clinicians evaluated PSNR-matched image pairs featuring different degradation types and selected the more clinically acceptable version in each case. These expert preferences were then compared against TinyUSFM-NRQ scores across 540 cross-degradation pairs to determine if the metric aligns with professional diagnostic standards when traditional pixel-fidelity measures fail to distinguish quality.

\subsection{Full-Reference Evaluation (TinyUSFM-uLPIPS)}
\label{sec:task_anchor_results}

\begin{figure}[t]
\centering
\includegraphics[width=\textwidth]{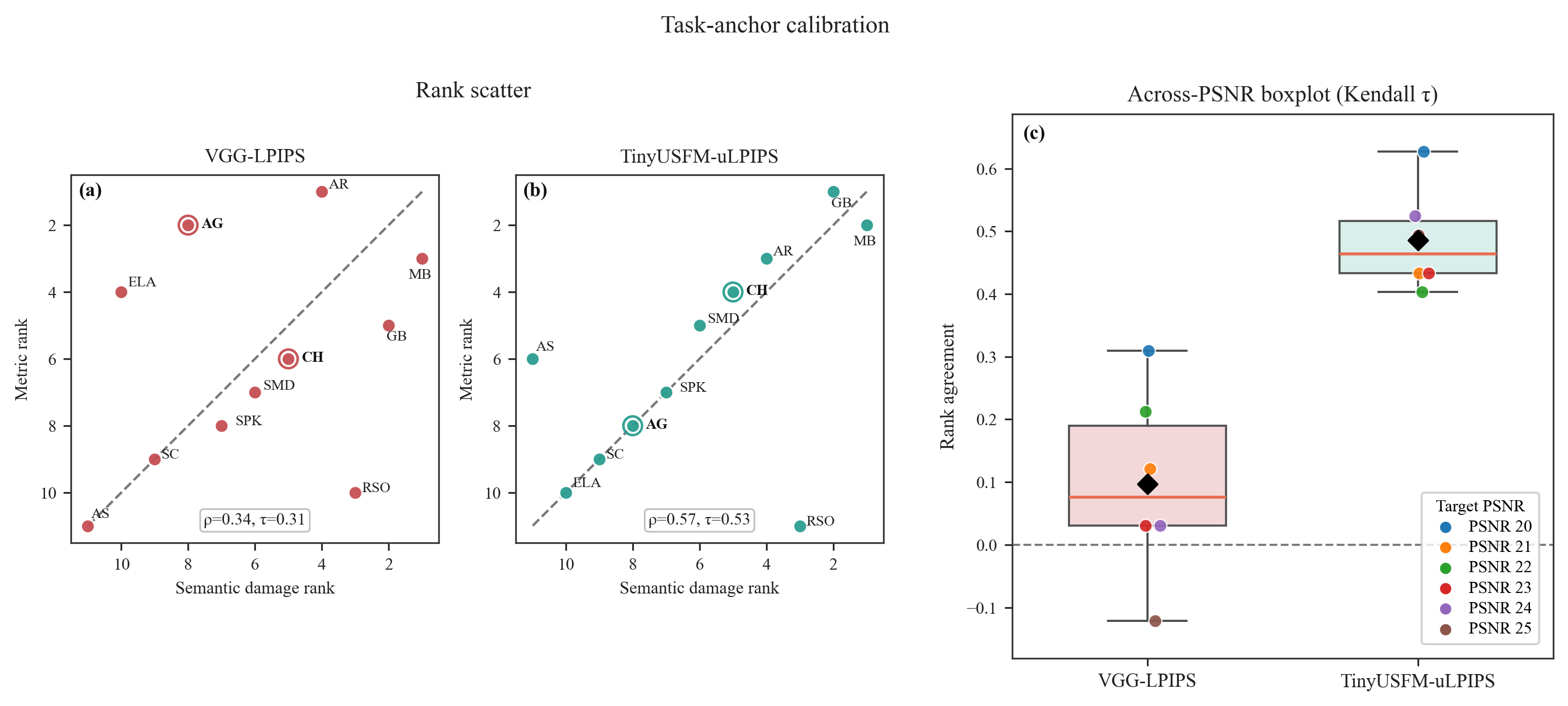}
\caption{Task-anchor calibration of FR metrics under PSNR-aligned evaluation. Left: representative DDTI rank--rank scatter at PSNR$=20$ for VGG-LPIPS and TinyUSFM-uLPIPS. Right: Kendall's $\tau$ summarized across segmentation anchors and PSNR targets; colored dots indicate individual anchor--PSNR settings.}
\label{fig:fr_rank_calib}
\end{figure}

\begin{figure}[t]
\centering
\includegraphics[width=\textwidth]{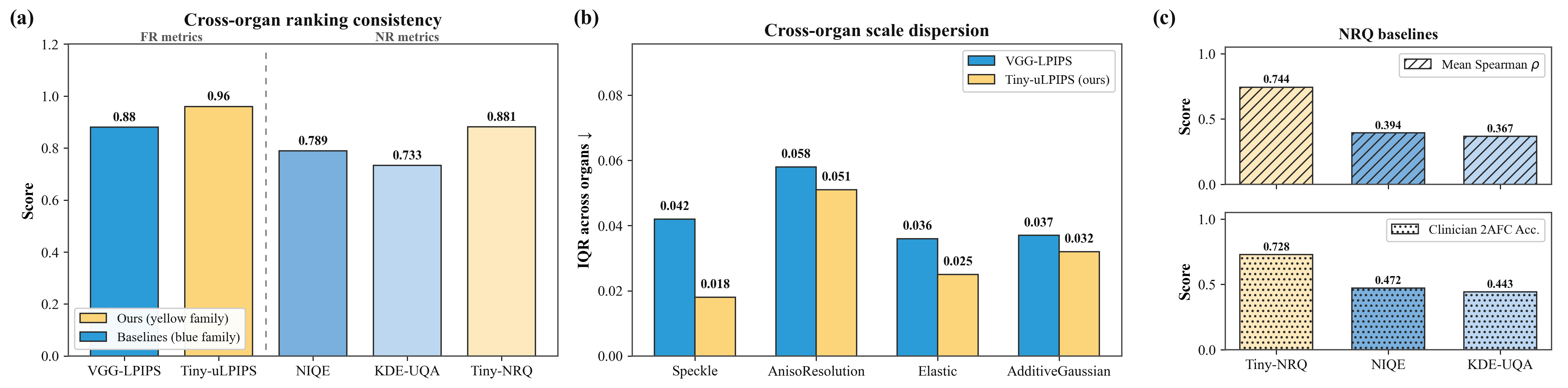}
\caption{Cross-organ robustness and NRQ baseline comparison under PSNR-aligned evaluation. (a) Cross-organ ranking consistency for FR and NRQ metrics (Kendall's $W$, PSNR 20--25 dB). (b) Cross-organ scale dispersion for FR metrics (IQR across organs) on representative degradations (aggregated over PSNR 20--25 dB). (c) NRQ baseline comparison: within-organ correlation (Spearman $\rho$) and clinician 2AFC agreement.}
\label{fig:baselines}
\end{figure}

\textbf{Experimental Setup:} We use thyroid and kidney ultrasound images for Dice-based segmentation evaluation, including thyroid nodule segmentation from DDTI \cite{pedraza2015ddti} and kidney capsule segmentation from the Open Kidney Ultrasound Data Set \cite{singla2023kidneyus_miccai}. We degrade the images to matched PSNR levels (20--25) and run the same segmentation model to obtain Dice scores. We then compare VGG-LPIPS and TinyUSFM-uLPIPS on the degraded images and test which metric better reflects the true severity of task damage.

\textbf{Task-Linked Quality:} Fig.~\ref{fig:fr_rank_calib} shows a representative DDTI rank-scatter example and Kendall's $\tau$ summaries across segmentation anchors and PSNR targets. TinyUSFM-uLPIPS aligns better with semantic damage than VGG-LPIPS on the thyroid and kidney ultrasound datasets, and it shows consistently higher and more stable Kendall's $\tau$ across PSNR targets (Fig.~\ref{fig:fr_rank_calib}). At low PSNR, the difference is most clear because distortion effects are easier to separate; at higher PSNR, rank comparison becomes harder because Dice drops are smaller, but TinyUSFM-uLPIPS remains more stable than VGG-LPIPS.

Under matched PSNR, VGG-LPIPS shows a systematic bias: it tends to over-penalize noise-like degradations and under-penalize ultrasound-specific structural artifacts. For example, at PSNR$=20$ on the thyroid dataset, VGG-LPIPS ranks AdditiveGaussian as more severe than ClutterHaze/ROIShadow Occlusion, even though the latter causes larger Dice drop, while TinyUSFM-uLPIPS gives the correct ordering.

\textbf{Cross-Organ Comparability:} TinyUSFM-uLPIPS produces more consistent cross-organ degradation rankings than VGG-LPIPS (Fig.~\ref{fig:baselines}(a)) and shows a smaller inter-organ IQR on several representative degradations, such as speckle-related and resolution-related corruptions (Fig.~\ref{fig:baselines}(b)).

\subsection{No-Reference Evaluation (TinyUSFM-NRQ)}
\label{sec:nrq_clinician}

\textbf{Experimental Setup.} We evaluate TinyUSFM-NRQ on public ultrasound datasets covering multiple organs and acquisition domains, including CUBS \cite{meiburger2022cubs}, UF1990 \cite{cai2024uf1990}, STMUS \cite{marzola2021stmusc}, AUL \cite{xu2023aul}, BUSI \cite{aldhabyani2020busi}, MMOTU \cite{zhao2022mmotu}, DDTI \cite{pedraza2015ddti}, KidneyUS \cite{singla2023kidneyus_miccai} and CAMUS \cite{leclerc2019camus}. Robustness analysis for TinyUSFM-NRQ and baselines is conducted on 9 organs following the shared protocol in Sec.~\ref{sec:datasets_preproc}, yielding (9$\times$8) organ$\times$distortion conditions. We compare TinyUSFM-NRQ with NIQE, BRISQUE, and KDE-UQA (Fig.~\ref{fig:baselines}(c)).

\textbf{PSNR-Consistent Sensitivity.} TinyUSFM-NRQ achieves a mean Spearman $\rho=0.744$ over PSNR levels, indicating strong within-organ monotonicity. NIQE and KDE-UQA fail systematically on ultrasound-typical structural degradations, especially scanline missing and specular clipping. At PSNR$=25$, these metrics become numerically ill-conditioned for structural artifacts, and several distortion families demonstrate near-zero or negative correlations with clinical quality across different organs. This makes them unreliable as a unified cross-organ quality control scale under heterogeneous ultrasound artifacts and is consistent with their lower monotonicity and weaker clinician alignment relative to TinyUSFM-NRQ (Fig.~\ref{fig:baselines}(c)).

Using the same PSNR-aligned degradation settings on unseen organs (stomach, multifidus muscle, and neck nerve), TinyUSFM-NRQ maintains strong monotonicity and reliable severity ranking under domain shift. Its degradation scoring trends are consistent with the in-distribution results, showing similar relative severity patterns across major artifact families.

\textbf{Clinician Utility.} TinyUSFM-NRQ predicts clinician preference with 72.8\% accuracy (95\% CI [0.689, 0.764], Wilson), significantly above chance (two-sided binomial test, $p{<}10^{-20}$). In embedded sanity checks (same degradation, different PSNR) and duplicate pairs, the readers achieved 100\% accuracy and consistency.

\subsection{Ultrasound Super-Resolution}
\label{sec:sr_results}

\textbf{Experimental Setup.} We train the same SwinIR backbone \cite{liang2021swinir} with identical splits under three objectives: $\mathcal{L}_1$, $\mathcal{L}_1+\mathcal{L}_{\mathrm{VGG}}$, and $\mathcal{L}_1+\mathcal{L}_{\mathrm{TinyUSFM}}$ (Sec.~\ref{sec:percep}). Training uses a combined breast+echocardiography ultrasound dataset for 100 epochs (LR $10^{-4}$) with $\lambda_{\mathrm{percep}}/\lambda_1=0.1/1$ for both perceptual losses. 

\begin{figure}[t!]
    \centering
    \includegraphics[width=\textwidth]{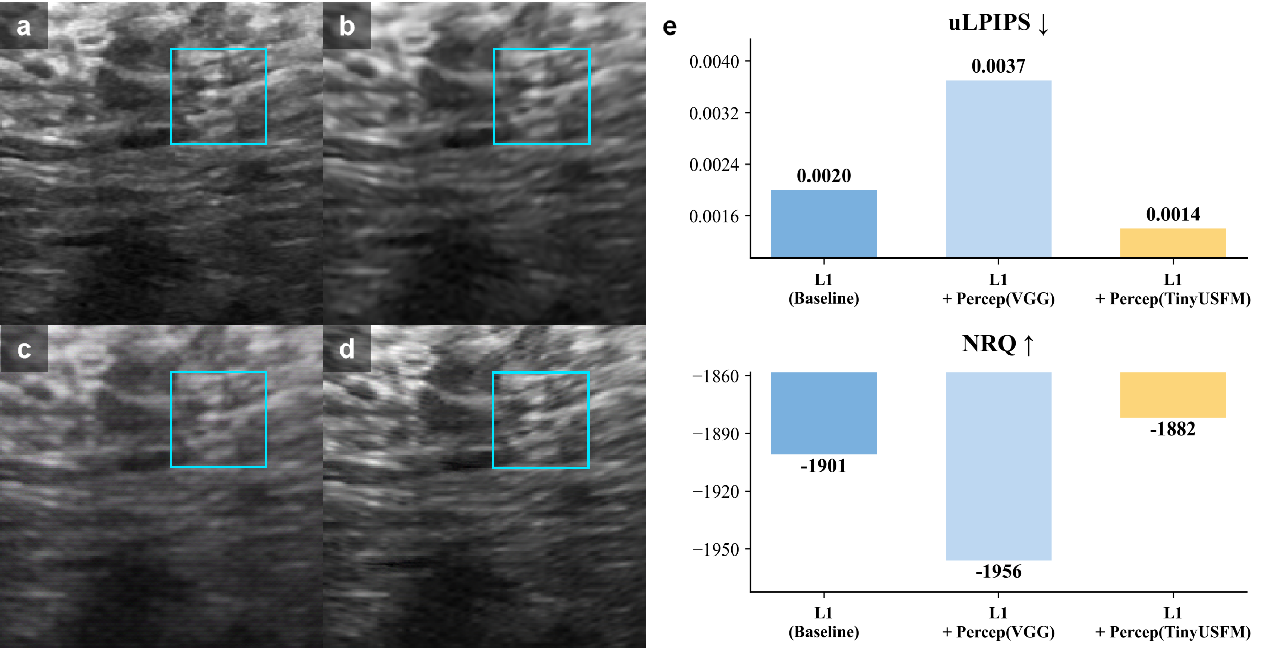}
    \caption{SR comparison under three training objectives. (a--d) Representative breast ultrasound example (ground truth; $\mathcal{L}_1$; $\mathcal{L}_1+\mathcal{L}_{\mathrm{VGG}}$; $\mathcal{L}_1+\mathcal{L}_{\mathrm{TinyUSFM}}$). (e) Quantitative comparison using TinyUSFM-uLPIPS and TinyUSFM-NRQ.}
    \label{fig:sr_results}
\end{figure}

\textbf{Perceptual-Clinical Alignment Loop.}
Compared with $\mathcal{L}_{\mathrm{VGG}}$, using $\mathcal{L}_{\mathrm{TinyUSFM}}$ as the perceptual loss improves PSNR and SSIM by 5.3\% and 3.3\%, respectively. Although $\mathcal{L}_1$ still gives the strongest PSNR, it produces over-smoothed reconstructions, while $\mathcal{L}_{\mathrm{VGG}}$ reduces fidelity and introduces ultrasound-inconsistent artifacts. In contrast, $\mathcal{L}_{\mathrm{TinyUSFM}}$ maintains competitive fidelity and better preserves diagnostically relevant acoustic texture with fewer conspicuous structural hallucinations. As shown in Fig.~\ref{fig:sr_results}, the ultrasound-native perceptual metrics (TinyUSFM-uLPIPS and TinyUSFM-NRQ) also favor $\mathcal{L}_{\mathrm{TinyUSFM}}$ reconstructions. These findings validate the use of ultrasound-native features for both model optimization and evaluation. By prioritizing acoustic-specific structures over pixel-level fidelity, the metric aligns more closely with clinical utility.

In a blinded clinician pairwise evaluation, expert sonographers compared reconstructions from the same case and selected the one more suitable for clinical diagnosis.
Clinicians preferred $\mathcal{L}_{\mathrm{TinyUSFM}}$ reconstructions over $\mathcal{L}_1$ in 87.5\% of cases and over $\mathcal{L}_{\mathrm{VGG}}$ in 100\% of cases.
Moreover, $\mathcal{L}_1$ was preferred over $\mathcal{L}_{\mathrm{VGG}}$ in 97.5\% of cases, indicating that visually sharper but hallucinated structures are clinically unacceptable in ultrasound.
This is consistent with TinyUSFM-NRQ, which favors $\mathcal{L}_{\mathrm{TinyUSFM}}$ reconstructions over the $\mathcal{L}_1$ and $\mathcal{L}_{\mathrm{VGG}}$ baselines.

Runtime analysis confirms a functional trade-off between the two modes. TinyUSFM-uLPIPS is optimized for high-fidelity offline research evaluation, whereas TinyUSFM-NRQ provides significantly lower latency and higher throughput. This efficiency makes the NRQ variant well-suited for real-time, no-reference quality control within active clinical and system pipelines.

\section{Conclusion}

We introduced a TinyUSFM-based ultrasound quality evaluation framework with two ultrasound-native metrics: TinyUSFM-uLPIPS for FR perceptual assessment and TinyUSFM-NRQ for NR quality scoring. Using a shared TinyUSFM feature space, the framework provides a modality-aligned alternative to pixel-fidelity metrics, natural-image perceptual metrics, and generic quality control baselines for ultrasound reconstruction and generative imaging. 
Future work will explore extending the framework to other ultrasound foundation models, higher-PSNR quality regimes, dynamic ultrasound, and diverse medical imaging modalities. The proposed metrics inherit the coverage and potential biases of the TinyUSFM backbone, and their reliability may be affected by rare pathologies, unseen devices, or under-represented acquisition settings. Overall, we believe this work provides the ultrasound community with a concrete answer to the question clinicians have always implicitly asked: Does this reconstruction show what I need to see?

\subsubsection{Acknowledgements}
This work was supported in part by the National Key R\&D Program of China
under Grant 2024YFF0507300 and Grant 2024YFF0507303, in part by the National
Natural Science Foundation of China under Grant 62531004, in part by the
Fudan University Medicine-Engineering Collaboration Program under Grant
yg-2025-general-07, and in part by the Shanghai Municipal Education Commission
Program on AI for Transforming Academic Research Paradigms (24RGZNB05).
\subsubsection{Disclosure of Interests}
The authors have no competing interests to declare.

\bibliographystyle{splncs04}
\bibliography{refs}
\end{document}